 \documentclass[final,number,3p,times,twocolumn]{elsarticle}

\usepackage{amssymb}

\usepackage{mathptmx}      
\usepackage{latexsym}
\usepackage{graphicx}
\usepackage{graphics}
\usepackage{epsfig}
\usepackage[latin1]{inputenc}
\usepackage{latexsym}
\usepackage{amsmath}
\usepackage{enumerate}
\usepackage{natbib}

\journal{Nuclear Instruments and Methods}

\begin{document}

\begin{frontmatter}



\title{Kochen-Specker theorem studied with neutron interferometer}


\author{Yuji Hasegawa \fnref{label2}}
\ead{Hasegawa@ati.ac.at}
\author{Katharina Durstberger-Rennhofer, Stephan Sponar, and Helmut Rauch}

\address{Atominstitut, Technische Universit\"{a}t Wien, Stadionallee 2, A-1020 Wien, Austria  }

\begin{abstract}
The Kochen-Specker theorem theoretically shows evidence of the incompatibility of noncontextual hidden variable theories with quantum mechanics. Quantum contextuality is a more general concept than quantum non-locality which is quite well tested in experiments by using Bell inequalities.
Within neutron interferometry we performed an experimental test  of the Kochen-Specker theorem with an inequality, which identifies quantum contextuality, by using spin-path entanglement in a single neutron system. Here entanglement is achieved not between different particles, but between degrees of freedom i.e., between spin and path degree of freedom. Appropriate combinations of the spin analysis and the position of the phase shifter allow an experimental verification of the violation of an inequality of the Kochen-Specker theorem. The observed violation $2.291\pm0.008 \nleq 1$ clearly shows that quantum mechanical predictions cannot be reproduced by noncontextual hidden variable theories.
\end{abstract}


\end{frontmatter}


\section{Introduction}

It was Einstein, Podolsky, and Rosen (EPR) \cite{EPR} and afterwards Bell \cite{Bell} who shed light on the non-local properties between subsystems in quantum mechanics. Separately Kochen and Specker \cite{KS67} analysed sets of measurements of compatible observables and found the impossibility of their consistent coexistence, i.e., quantum indefiniteness of measurement results. In their scenario, quantum contextuality, a more general concept compared to non-locality, leads to striking phenomena predicted by quantum theory.

Bell inequalities \cite{Bell} are constraints imposed by local hidden-variable theories (LHVTs) on the values of some specific linear combinations of the averages of the results of spacelike separated experiments on distant systems. Reported experimental violations of Bell inequalities, e.g. with photons \cite{WJSWZ98}, neutrons \cite{Hasegawa03} or atoms \cite{Matsukevich}, suggest that quantum mechanics (QM) cannot be reproduced by LHVTs.

 While violations of Bell's inequalities due to nonlocal characters of QM is impressive, conflict between measurements on a single-system is another marvelous prediction of QM, as is first stated by Kochen-Specker \cite{KS67}. Quantum mechanical peculiarity is not limited to spacelike separated systems, but
found in measurements of a single nonsepared system: it is important to investigate the consequences of hidden-variable theories for (massive) non-spacelike separated quantum systems, such as neutrons.


LHVTs are a subset of a larger class of hidden-variable theories known as noncontextual hidden-variable theories (NCHVTs). In NCHVTs the result of a measurement of an observable is assumed to be predetermined and not affected by a (previous or simultaneous) suitable measurement of any other compatible or co-measurable observable. It turns out that there exists a conflict between the predictions of QM and NCHVTs which is predicted by the KS theorem \cite{KS67}.

Here, we describe experimental demonstration of the violation in line with the KS theorem by using a massive quantum systems, in particluar, two degrees of freedom of single neutrons within a neutron interferometer.

\section{Kochen-Specker theorem}

The Kochen-Specker (KS) theorem states that there is no contextual hidden variable model possible that reproduces the predictions of QM (for a review see, e.g., Ref. \cite{Mermin93}).

The theorem uses two assumptions: (a) value definiteness (all observables defined for a system, e.g. $A$ and $B$, have predefined values, e.g. $v(A)$ and $v(B)$) and (b) noncontextuality (a system possesses a property independently of any measurement context, i.e., independently of how the value is measured).
Due to assumption of noncontextuality the relations $v(A+B)=v(A)+v(B)$ and $v(A\cdot B)=v(A) \cdot v(B)$ hold for mutually compatible observables, which have a set of common eigenvectors and thus are together measureable. One can show mathematically that it is impossible to satisfy both relations  for arbitrary pairs of compatible operators $A$ and $B$ within QM.

The original proof was given by Kochen and Specker \cite{KS67} in 1967 which involves 117 vectors in 3 dimensions. Later on simpler proofs were found, e.g., for 9 observables in 4 dimensions (2 spin-$\frac{1}{2}$ particles) by Peres \cite{Peres90} and Mermin \cite{Mermin93} who extended Peres' proof into a state independent proof, and with 10 observables in 8 dimensions (3 spin-$\frac{1}{2}$ particles) by Mermin \cite{Mermin90} who could also show a connection to the Greenberger-Horne-Zeilinger (GHZ) version \cite{GHZ} of Bells theorem. Up to now the simplest proof of the KS theorem was found by Cabello \cite{CEG96} which uses 18 vectors in 4 dimensions.

We give a short explanation of the proof by Peres and Mermin discussed in Ref. \cite{Mermin93}. In 4 dimensions observables are represented by Pauli matrices of two spin-$\frac{1}{2}$ particles $\sigma_i^{1}$ and $\sigma_j^{2}$ where $i,j=\{x,y,z\}$. The square of each matrix is unity, the eigenvalues are $\pm1$, in each subspace the normal commutation relations for Pauli operators are satisfied, and the commutator of observables from different subspaces is zero $[\sigma_i^{1},\sigma_j^{2}]=0$ for any $i,j$.
Consider the following 9 observables $A_m$ arranged in a ``magic square'':
\begin{align*}
&\;\;\sigma_x^1 &         &\;\;\sigma_x^2 &       \sigma_x^1 &\cdot \sigma_x^2   &    &\longrightarrow +1\\
&\;\;\sigma_y^2 &         &\;\;\sigma_y^1 &       \sigma_y^1 &\cdot \sigma_y^2   &    &\longrightarrow +1\\
&\sigma_x^1\sigma_y^2   &     &\sigma_y^1\sigma_x^2   &    \sigma_z^1 &\cdot \sigma_z^2
& &\longrightarrow +1\\
&\;\;\downarrow    &         &\;\;\downarrow   &       &\downarrow  & &\\
&\;+1   &         &\;+1    &       &-1     & &
\end{align*}
In each row and column the observables are mutually commuting and hence compatible. The product of three observables in each row and also in the first two columns is $+1$ but in the last column we get $-1$ for the product (due to $\sigma_x^k\sigma_y^k=i\sigma_z^k$ for $k=1,2$). Thus the product of all rows and colums is $-1$. In NCHVTs we assign to each observable a definite value $v(A_m)$. If we take the product over all rows and colums each value $v(A_m)$ appears twice leading to a total product of $+1$. This contradicts the QM predictions.

In contrast to Bell's theorem the KS theorem does not use statistical predictions but relies on logical contradictions. However, in experiments we have finite precession and thus never perfect correlations which makes it necessary to deduce experimentally testabel inequalities from the KS theorem. There are proposals which use the assumptions of contextuality but are not independent from additional QM predictions, e.g. \cite{SimonBruknerZeilinger2001}, but there are also inequalities which are based only on the assumptions of contextuality \cite{Cabello08a}. There exist also state-independet inequalities to test KS theorem \cite{Cabello08b} and the first experiments were done with single photons \cite{Amselem09} and ions \cite{Kirchmair09} confirming a violation of the inequality.

\begin{figure}[h!]
\begin{center}
\includegraphics[width=0.39\textwidth]{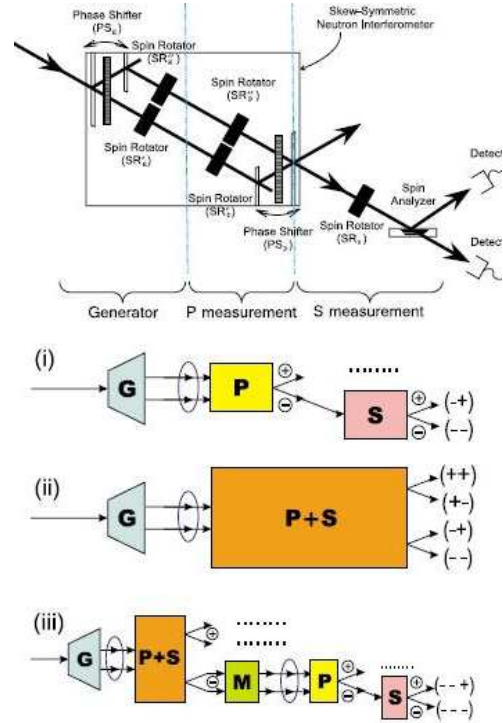}
\caption{ Above: A proposed experimental setup with a
neutron interferometer. The interferometer is set in a way that fulfills
two functions: the first half works as a state generator, and the second half works as a path
measurement apparatus. In both parts, a phase shifter (PS) as well
as a pair of spin rotators (SR) are inserted. A spin measurement is carried out
on the outgoing beam in the forward direction.
Below: Three diagrams for the different measurement "contexts". (i) For
measurements of $\sigma_x^s \cdot \sigma_x^p$ and $\sigma_y^s
\cdot \sigma_y^p$: After going through a state generator (G), a
state suffers a path measurement (P) followed by a spin
measurement (S). Consequently, each outgoing beam gives the
results of the two measurements. (ii) For measurements of
$\sigma_y^s \sigma_x^p\cdot\sigma_x^s \sigma_y^p$: By tuning one
of the spin rotators to a spin-flip operation in the path
measurement part, the second half of the interferometer together
with a spin analyzer (P+S) can discriminate four Bell states, which
assign four outgoing beams to the four possible results of
the measurements. (iii) For measurements of $\sigma_y^s \sigma_x^p
\cdot \sigma_y^s \cdot \sigma_x^p$ and $\sigma_x^s \sigma_y^p
\cdot \sigma_x^s \cdot \sigma_y^p$: After the apparatus P+S, a
state mixer (M) eliminates the former information about the result
of either observable, and is followed by a path and a spin
measurement.}
\label{Fig:1}
\end{center}
\end{figure}

\section{Theoretical considerations}

With the use of an inequality of the KS theorem \cite{Cabello08a} one can study the statistical violation of non-contextual assumptions.
Exploiting the interference effect of matter waves together with entanglement in a single-particle system, neutron interferometric experiments \cite{Rauch00} are suitable to exhibit phenomena associated with the KS theorem. At the first stage of experimental tests of quantum contextuality,
we carried out interferometric experiments demonstrating Kochen-Specker-like phenomena \cite{Hasegawa06}. Further theoretical analysis revealed a more advanced scheme  based
on the Peres-Mermin proof of the KS theorem and an experiment with neutron interferometry was proposed \cite{Cabello08a}. Here, an improved experimental test of
the KS theorem using single neutrons is described where the entanglement occurs between two degrees of freedom in a single-system \cite{Bartosik09}.

\begin{figure*}
\begin{center}
   \includegraphics[width=0.60\textwidth]{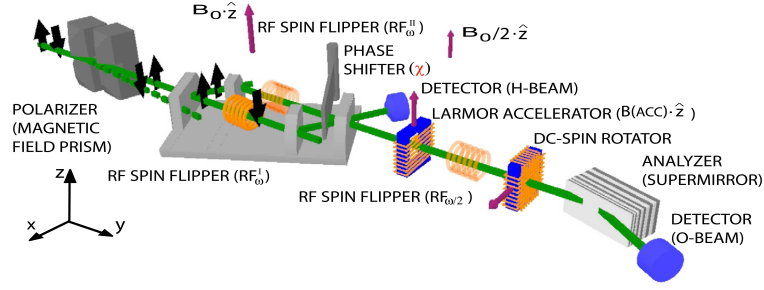}
   \caption{Experimental setup for studying Kochen-Specker theorem based on the Peres-Mermin proof with neutron interferometer. The RF flipper in the path I  (RF$_{\omega}^I$) generates the Bell-like state $|\Psi^{Bell}_n \rangle$. By turning either the RF flipper in the path II (RF$_{\omega}^{II}$) or another RF flipper (RF$_{\omega/2}$) on, together with suitable spin analysis, intensity oscillations are obtained in phase shifter $\chi$ scans. From the data on the appropriate settings, expectation values of the measurements
   $\sigma ^{s} _{x} \cdot \sigma ^{p} _{x}, \sigma ^{s} _{y} \cdot \sigma ^{p} _{y}$, and
   $\sigma^{s} _{x} \sigma ^{p} _{y} \cdot \sigma ^{s} _{y} \sigma ^{p} _{x}$ are determined. }
\label{fig:2} \end{center}
\end{figure*}

For the proof of the KS theorem, we consider single neutrons prepared in a maximally entangled Bell-like state
\begin{equation}\label{Bell-like-entangled-state}
|\Psi^{Bell}_n \rangle=\frac{1}{\sqrt2}(| \downarrow \rangle\otimes | {\rm I} \rangle -
		| \uparrow \rangle\otimes | {\rm II} \rangle),
\end{equation}
where $| \uparrow \rangle$ and $| \downarrow \rangle$ denote spin-up and spin-down eigenstates of the neutron, and $| {\rm I} \rangle$ and $| {\rm II} \rangle$ denote the two beam paths in the neutron interferometer \cite{Hasegawa03}.
The proof is based on six observables $\sigma_{x}^{s}$, $\sigma_{x}^{p}$, $\sigma_{y}^{s}$, $\sigma_{y}^{p}$, $\sigma_{x}^{s}\sigma_{y}^{p}$ and $\sigma_{y}^{s}\sigma_{x}^{p}$, where the superscripts $s$ and $p$ indicate the spin and path degree of freedom, respectively, and the following five quantum mechanical predictions for the Bell-like state $|\Psi^{Bell}_n\rangle$
\begin{subequations}
\begin{align}
\sigma ^{s} _{x} \cdot \sigma ^{p} _{x} \:| \Psi^{Bell}_n \rangle &=  -| \Psi^{Bell}_n \rangle,  \label{qmpredictionsa}
\\
\sigma ^{s} _{y} \cdot \sigma ^{p} _{y} \:| \Psi^{Bell}_n \rangle &=  -| \Psi^{Bell}_n \rangle, \label{qmpredictionsb}
\\
\sigma ^{s} _{x} \sigma ^{p} _{y} \cdot \sigma ^{s} _{x} \cdot \sigma ^{p} _{y} \:| \Psi^{Bell}_n \rangle & = + | \Psi^{Bell}_n \rangle, \label{qmpredictionsc}
\\
\sigma ^{s} _{y} \sigma ^{p} _{x} \cdot \sigma ^{s} _{y} \cdot \sigma ^{p} _{x} \:| \Psi^{Bell}_n \rangle &= +| \Psi^{Bell}_n \rangle, \label{qmpredictionsd}
\\
\sigma ^{s} _{x} \sigma ^{p} _{y} \cdot \sigma ^{s} _{y} \sigma ^{p} _{x} \:| \Psi^{Bell}_n \rangle &= -| \Psi^{Bell}_n \rangle. \label{qmpredictionse}
\end{align}
\end{subequations}

The inconsistency arising in any attempt to ascribe the predefined values $-1$ or $+1$ to each and every of the six observables can be easily seen by multiplying
Eqs.~(\ref{qmpredictionsa})-(\ref{qmpredictionse}). Since each observable appears twice, the left hand sides give $+1$ while the product of the right hand sides is $-1$.

Since no experiment can show perfect correlations or anti-correlations, one needs an experimentally testable inequality: this can be derived from the linear combination of the five expectation values with the respective quantum mechanical predictions as linear coefficients. It can be shown that in any NCHVT
\begin{eqnarray}
-\langle \sigma ^{s} _{x} \cdot \sigma ^{p} _{x} \rangle - \langle \sigma ^{s} _{y} \cdot \sigma ^{p} _{y} \rangle + \langle \sigma ^{s} _{x} \sigma ^{p} _{y} \cdot \sigma ^{s} _{x} \cdot \sigma ^{p} _{y} \rangle  \hfill\qquad\qquad\:\nonumber\\
+ \:\langle \sigma ^{s} _{y} \sigma ^{p} _{x} \cdot \sigma ^{s} _{y} \cdot \sigma ^{p} _{x} \rangle - \langle \sigma ^{s} _{x} \sigma ^{p} _{y} \cdot \sigma ^{s} _{y} \sigma ^{p} _{x} \rangle \leq 3, \label{inequality3}
\end{eqnarray}
in contrast to the prediction of 5 by QM. While Eqs.~(\ref{qmpredictionsa})-(\ref{qmpredictionsb}) and (\ref{qmpredictionse}) represent state
dependent predictions relying on the specific properties of the neutron's Bell-like state $|\Psi^{Bell}_n\rangle$, Eqs.~(\ref{qmpredictionsc})-(\ref{qmpredictionsd}) are
state-independent predictions which hold in any NCHVT. In other words, in any NCHVT, $\langle \sigma ^{s} _{x} \sigma ^{p} _{y} \cdot \sigma
^{s} _{x} \cdot \sigma ^{p} _{y} \rangle = 1$ and $\langle \sigma ^{s} _{y} \sigma ^{p} _{x} \cdot \sigma ^{s} _{y} \cdot \sigma ^{p} _{x}
\rangle =1$. Therefore, any NCHVT must satisfy not only inequality~(\ref{inequality3}), but also the following inequality in a reduced form:
\begin{equation}
-\langle \sigma ^{s} _{x} \cdot \sigma ^{p} _{x} \rangle - \langle
\sigma ^{s} _{y} \cdot \sigma ^{p} _{y} \rangle  - \langle \sigma
^{s} _{x} \sigma ^{p} _{y} \cdot \sigma ^{s} _{y} \sigma ^{p} _{x}
\rangle \leq 1. \label{inequtbtested}
\end{equation}
A violation of inequality~(\ref{inequtbtested}) in an experiment reveals quantum contextuality.

\section{Neutron interferometric experiments}

The experiment was carried out at the neutron interferometer
instrument S18 at the high-flux reactor of the Institute
Laue-Langevin (ILL) in Grenoble, France. The setup of the experiment
is depicted in Fig.{\ref{fig:2}}. A monochromatic beam, with mean
wavelength $\lambda_0=1.92 \mbox{ \AA}
(\Delta\lambda/\lambda_0\sim0.02$) and 5x5 mm$^2$ beam
cross-section, is polarized by a bi-refringent magnetic field prism
in $\hat{\mathbf z}$-direction. Due to the angular separation at the
deflection, the interferometer is adjusted so that only the spin-up
component fulfills the Bragg condition at the first interferometer
plate (beam splitter). Behind the beam splitter  the neutrons wave
function is found in a coherent superposition of path $| I \rangle$
and $| II \rangle$. Together with a radio-frequency (RF)
spin-flipper in path $| I \rangle$, denoted as RF$_{\omega}^I$, the
first half of the interferometer is used for the generation of the
maximally entangled Bell-like state, Eq.~(\ref{Bell-like-entangled-state}).
In this experiment, RF spin-flippers are used for the spin-flips to
avoid unwanted contrast reduction due to dephasing effect by the
Mu-metal, used in the previous experiment \cite{Hasegawa06}.
Apart from the RF flipper in path $| I \rangle$ our
experiment requires a second RF flipper in the interferometer
(RF$_{\omega}^{I\!I}$) and another RF flipper in the O-beam
(in the forward direction) operated
at half frequency (RF$_{\omega /2}$).

The first term in inequality (\ref{inequtbtested}) requires the
measurement of  $\sigma ^{s} _{x}$ together with $\sigma ^{p} _{x}$.
Here, RF$_{\omega /2}$ in the O-beam is needed for compensating the
energy difference due to the spin flip at RF$_{\omega}^I$
\cite{Sponar08a}, while the second RF flipper in the interferometer,
RF$_{\omega}^{I\!I}$, is turned off. For measuring the path
observable, i.e. $\sigma ^{p} _{x}$, the phase shifter is adjusted
to $ \chi = 0 $ and $ \chi = \pi $ in the path state
$|\Psi(\chi)\rangle_{p} = \frac{1}{\sqrt{2}} (| I \rangle + e^{i
\chi} | II \rangle)$, which correspond to the projections to
$|+x\rangle_{p}$ and $|-x\rangle_{p}$, the two eigenstates of
$\sigma ^{p} _{x}$, respectively. The spin analysis in the $x-y$
plane is accomplished by the combination of the Larmor accelerator
DC coil inducing a Larmor phase $\alpha = 0$ and $\alpha = \pi$, a
$\pi/2$ DC spin-rotator and  an analyzing supermirror. This
configuration allows projective measurements along $|+x\rangle_{s}$
and $|-x\rangle_{s}$ direction, the two eigenstates of $\sigma ^{s}
_{x}$.

\begin{figure}
\begin{center}
   \includegraphics[width=0.3\textwidth]{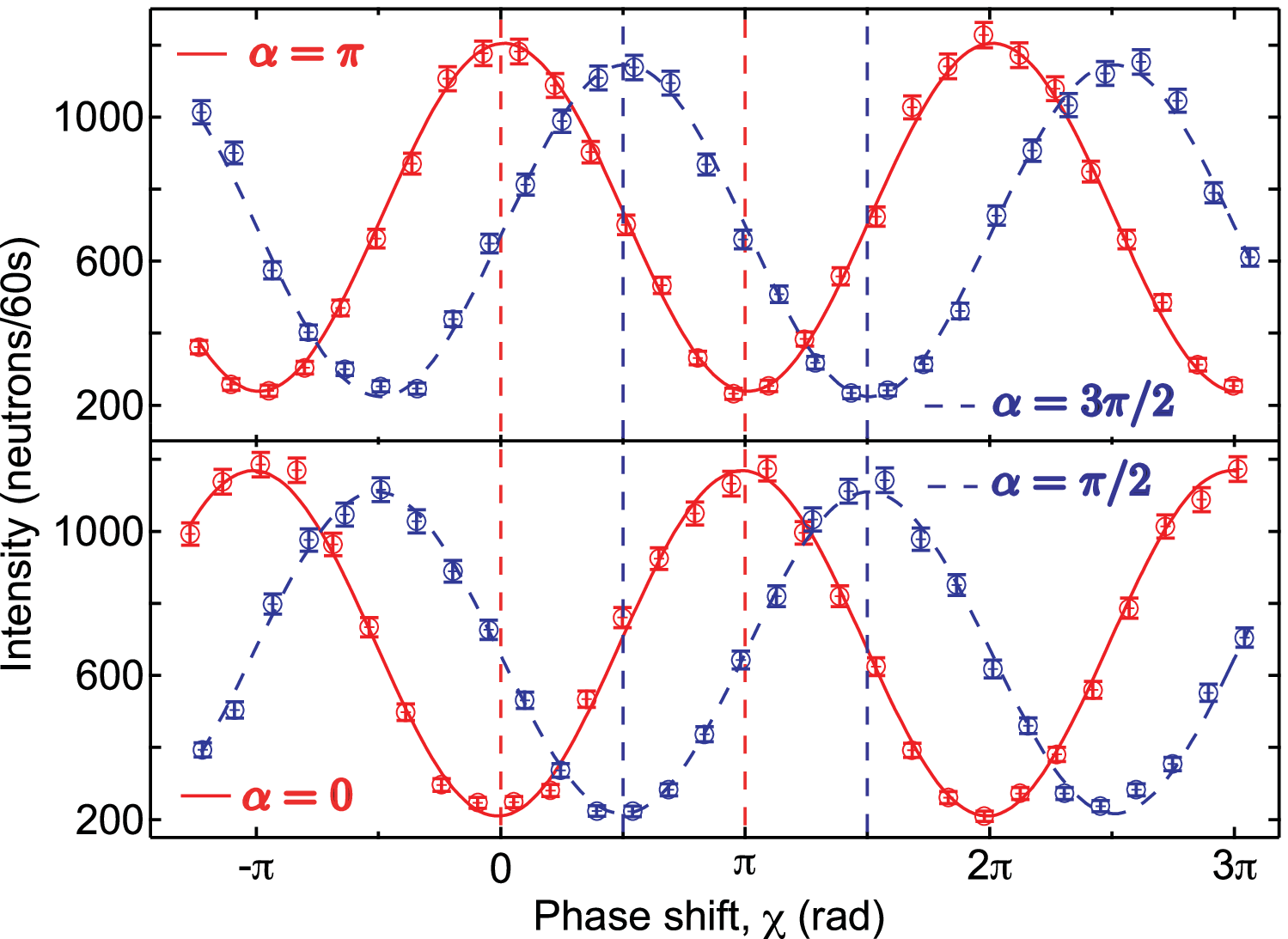}
   \hspace{10pt}
   \includegraphics[width=0.3\textwidth]{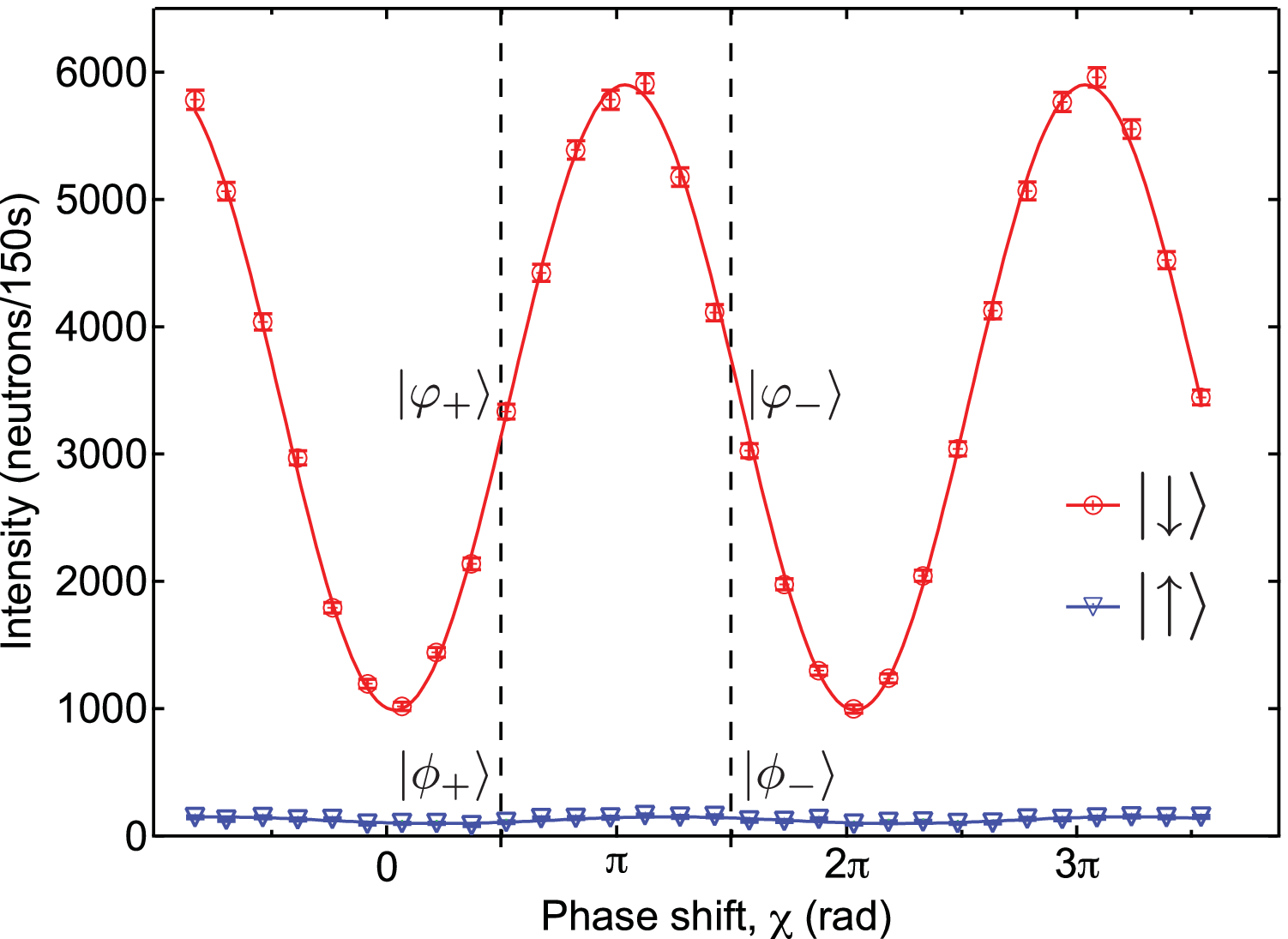}
   \caption{Typical intensity modulations obtained by varying the phase $\chi$ for the path subspace.
   The spin analysis of $\pm x-$ and $\pm y-$directions were involved (top). Another spin-flipper in the interferometer
   was turned on and the spin analysis of $\pm z-$ directions were carried out (bottom).  }
\label{fig:3}       
\end{center}
\end{figure}

The experimental setup for the second term in inequality
(\ref{inequtbtested}) is identical with the one for the first term,
but the measurement of $\sigma ^{s} _{y}$ together with $\sigma ^{p}
_{y}$ is achieved with the settings $\chi = \frac{\pi}{2},
\frac{3\pi}{2}$ and $\alpha = \frac{\pi}{2} , \frac{3\pi}{2}$.
Typical intensity oscillations for the successive measurement of the
path and the spin component are shown in Fig.~\ref{fig:3} top. The
expectation values are experimentally determined from the count
rates
\begin{equation}
E(\alpha, \chi) = \tfrac{N(\alpha, \chi) + N(\alpha + \pi, \chi +
\pi) - N(\alpha + \pi, \chi) - N(\alpha, \chi + \pi)}{N(\alpha,
\chi) + N(\alpha + \pi, \chi + \pi) + N(\alpha + \pi, \chi) +
N(\alpha, \chi + \pi)}\label{equ:expvalue},
\end{equation}
where $N(\alpha,\chi)$ denotes the count rate for the joint
measurement of spin and path. The required count rates at
appropriate settings of $\alpha$ and $\chi$ are extracted from least
squares fits in Fig.~\ref{fig:3} top, indicated by the vertical
dashed lines. From these intensities  the expectation values were
determined as $\langle \sigma ^{s} _{x} \cdot \sigma ^{p} _{x}
\rangle \equiv E(0, 0) = -0.679 \pm 0.005$ and $\langle \sigma ^{s}
_{y} \cdot \sigma ^{p} _{y} \rangle \equiv E(\frac{\pi}{2},
\frac{\pi}{2})= -0.682 \pm 0.005$.

The third term in inequality (\ref{inequtbtested}) requires the
measurement of $\sigma ^{s} _{x} \sigma ^{p} _{y}$ together with
$\sigma ^{s} _{y} \sigma ^{p} _{x}$. Measuring the product of these
two observables simultaneously implies the discrimination of the
four possible outcomes $(\sigma ^{s} _{x} \sigma ^{p} _{y} , \sigma
^{s} _{y} \sigma ^{p} _{x}) =
\left\{(+1,+1),(-1,-1),(+1,-1),(-1,+1)\right\},$ which is equivalent
to a complete Bell-state discrimination. The two operators $\sigma
^{s} _{x} \sigma ^{p} _{y}$ and $\sigma ^{s} _{y} \sigma ^{p} _{x}$
have the four common Bell-like eigenstates
\begin{subequations}
\begin{align}
| \varphi _{\pm} \rangle &= \tfrac{1}{\sqrt{2}} ( \mid \downarrow
\rangle \otimes | I \rangle \pm i \mid \uparrow \rangle \otimes | II
\rangle),\label{def:varphi+-}
\\
| \phi _{\pm} \rangle &= \tfrac{1}{\sqrt{2}} ( \mid \uparrow \rangle
\otimes | I \rangle \pm i \mid \downarrow \rangle \otimes | II
\rangle),\label{def:phi+-}
\end{align}
\end{subequations}
with the corresponding eigenvalue equations
\begin{subequations}
\begin{align}
\sigma ^{s} _{x} \sigma ^{p} _{y} ~| \varphi _{\pm} \rangle &= \pm |
\varphi _{\pm} \rangle, &\sigma ^{s} _{y} \sigma ^{p} _{x} ~|
\varphi _{\pm} \rangle &= \mp | \varphi _{\pm} \rangle,
\label{xsyp-ysxpEW1}
\\
\sigma ^{s} _{x} \sigma ^{p} _{y} ~| \phi _{\pm} \rangle &= \pm |
\phi _{\pm} \rangle, & \sigma ^{s} _{y} \sigma ^{p} _{x} ~|
\phi _{\pm} \rangle &= \pm | \phi _{\pm} \rangle.
\label{ysxp-xsypEW2}
\end{align}
\end{subequations}

It follows that the outcome $-1$ for the product measurement of
$\sigma ^{s} _{x} \sigma ^{p} _{y}$ and $\sigma ^{s} _{y} \sigma
^{p} _{x}$ is obtained for $| \varphi _{\pm} \rangle$, while the
states $| \phi _{\pm} \rangle$ yield the result $+1$. In practice,
this Bell-state discrimination is accomplished by the second RF
flipper in the interferometer i.e.~transforming the state $| \Psi
\rangle \rightarrow \frac{1}{\sqrt{2}} (\mid \downarrow \rangle
\otimes | I \rangle - \mid \downarrow \rangle  \otimes | II
\rangle)$. When the DC spin-rotator in the O-beam is adjusted to
induce a $\pi$-flip, only $\mid\downarrow \rangle$-spin components
reach the detector. Inducing a relative phase $\chi$ between the two
beam paths in the interferometer allows then for projections to the
state $| \varphi (\chi)\rangle = \frac{1}{\sqrt{2}}(\mid \downarrow
\rangle \otimes | I \rangle + e^{i \chi} \mid \uparrow \rangle
\otimes | II \rangle)$. According to the definition of $| \varphi
_{\pm} \rangle$, given in Eq.(\ref{def:varphi+-}), phase settings of
$\chi = \pm \frac{\pi}{2}$ correspond to the measurement of $|
\varphi _{\pm} \rangle$. The $\mid\uparrow \rangle$-spin analysis is
achieved by switching the DC spin-rotator off, where neutrons in the
state $| \phi (\chi)\rangle = \frac{1}{\sqrt{2}}(\mid \uparrow
\rangle \otimes | I \rangle + e^{i \chi} \mid \downarrow \rangle
\otimes | II \rangle)$ can be selected, yielding a $| \phi _{\pm}
\rangle$ measurement for $\chi = \pm \frac{\pi}{2}$. By rotating the
phase shifter, clear sinusoidal intensity oscillation and a
low-intensity fluctuation were observed, which is depicted in
Fig.{\ref{fig:3} bottom. The expectation value $\langle \sigma ^{s}
_{x} \sigma ^{p} _{y} \cdot \sigma ^{s} _{y} \sigma ^{p} _{x}
\rangle$ is derived using the relation
\begin{equation}
E' = \tfrac{N'(\phi _{+}) + N'(\phi _{-}) - N'(\varphi _{+}) - N'(
\varphi _{-})}{N'(\phi _{+})+ N'(\phi _{-}) + N'(\varphi _{+}) +
N'(\varphi _{-})}\label{equ:expvaluebelldiscr},
\end{equation}
where $N'$ denotes the neutron count rate at the desired
projections. As done before, least square fits were applied to
deduce the count rates at the four projections. From the intensities
on the dashed lines in the figure, we obtained the value $\langle
\sigma ^{s} _{x} \sigma ^{p} _{y} \cdot \sigma ^{s} _{y} \sigma ^{p}
_{x} \rangle \equiv E' = -0.93 \pm 0.003$. The observed intensities
reflect the quantum mechanical predictions for the measurement of
the four Bell-like states given by the expectation values $\langle
\Psi | \varphi _{\pm} \rangle \langle \varphi _{\pm} | \Psi \rangle
= \frac{1}{2}$ and $\langle \Psi | \phi _{\pm} \rangle \langle \phi
_{\pm} | \Psi \rangle = 0$.

With the three experimentally derived expectation values  we can
finally test inequality (\ref{inequtbtested}). We obtain $-\langle
\sigma ^{s} _{x} \cdot \sigma ^{p} _{x} \rangle -\langle \sigma ^{s}
_{y} \cdot \sigma ^{p} _{y} \rangle
 - \langle \sigma ^{s} _{x} \sigma ^{p} _{y} \cdot \sigma ^{s} _{y} \sigma ^{p} _{x} \rangle = 2.291 \pm 0.008 \not\leq 1$.
This clearly confirms the conflict with NCHVTs.

\section{Concluding remarks}

Neutron interferometric investigations on the KS theorem is described.
Entanglement between degrees of freedom in a single-neutron system is exploited:
a Bell-like state comprising spin-path entanglement is generated. The proof is based on
the Peres-Mermin criteria. An inequality was derived for the evaluation of the experimental data.
Expectation values of three different contexts are determined: the final results
clearly exhibit the conflict between NCHVTs and QM. It is worth mentioning that
the conflict of the KS theorem is not due to the entanglement but can be assigned to
the structure of observables: quantum observables have a particular structure which is
different from that of classical observables. We now proceed further studies
of quantum contextuality with the use of triply entangled (spin-path-energy entangled) states
in a single-neutron system.  In addition, neutron polarimeter experiments are used for similar studies,
where tunable multi energy levels in addition to spin can be manipulated with very high efficiency.



\section*{Acknowledgements}
We thank all colleagues who were involved in carrying out experiments presented here;
in particular, we appreciate G. Badurek, H. Bartosik, A. Cabello, S. Filipp, D. Home,
J. Klepp, R. Loidl, and C. Schmitzer.
This work has been supported partly by the Austrian Fonds zur F\"{o}derung der Wissenschaftlichen Forschung (FWF), No. P21193-N20 and Hertha-Firnberg-Programm T389-N16.



\bibliographystyle{elsarticle-harv}



\end{document}